\documentclass[pra,aps,showpacs,twocolumn,superscriptaddress]{revtex4-1}

\usepackage{times}
\usepackage{amssymb}
\usepackage{amsmath}
\usepackage{mathrsfs}
\usepackage{graphicx}
\usepackage{epsfig}
\usepackage{dcolumn}
\usepackage{color}
\usepackage{bm}
\usepackage{cleveref}

\usepackage[normalem]{ulem}

\newcommand{\figref}[1]{Figure~\ref{#1}}

\newcommand{\eqnref}[1]{Equation~(\ref{#1})}

\begin{document}

\author{Vincent E. Elfving*}
\email{vincent.elfving@quandco.com}
\affiliation{Qu \& Co B.V., 1070 AW Amsterdam, The Netherlands}
\author{Marta Millaruelo}
\author{Jos\'e A. G\'amez}
\email{jose.gamez@covestro.com}
\author{Christian Gogolin}
\affiliation{Covestro Deutschland AG, 51365 Leverkusen, Germany}

\title{Simulating quantum chemistry in the seniority-zero space on qubit-based quantum computers}
\date{\today}

%******************************************************************************************************************************************************************************
%*****************************************************************************************************************************************************************************
\begin{abstract}
	Accurate quantum chemistry simulations remain challenging on classical computers for problems of industrially relevant sizes and there is reason for hope that quantum computing may help push the boundaries of what is technically feasible.
	While variational quantum eigensolver (VQE) algorithms may already turn noisy intermediate scale quantum (NISQ) devices into useful machines, one has to make all efforts to use the scarce quantum resources as efficiently as possible.
	We combine the so-called seniority-zero, or paired-electron, approximation of computational quantum chemistry with techniques for simulating molecular chemistry on gate-based quantum computers and obtain a very resource efficient quantum simulation algorithm.
	While some accuracy is lost through the paired-electron approximation, we show that using the freed-up quantum resources for increasing the basis set can lead to more accurate results and reductions in the necessary number of quantum computing runs by several orders of magnitude, already for a simple system like lithium hydride. We also discuss an error mitigation scheme based on post-selection which shows an attractive scaling when the given Hamiltonian format is considered, increasing the viability of its NISQ implementation.
\end{abstract}

\maketitle

%******************************************************************************************************************************************************************************
%*****************************************************************************************************************************************************************************

\section{Introduction}
One of the most promising near term applications of quantum computers is the simulation of quantum Hamiltonians in quantum chemistry. 
Current classical algorithms for accurately simulating the behavior of molecules are computationally very costly.
These can be classified into two main categories: based on the electronic wavefunction of the molecular system or on their electronic density.
In the latter case, known as density functional theory (DFT), the accuracy of the results depends on the system investigated, which renders them not universally applicable and makes necessary a prior assessment.
Among the wave function based methods, two are noteworthy because of their universally high accuracy: Full-Configuration Interaction (FCI) and Coupled Cluster (CC) theory.
The FCI solution of an electronic structure Hamiltonian is numerically exact but requires exponential time on a classical computer, limiting its applicability to systems with very few atoms and/or electrons.
Within the CC techniques, the CCSD (coupled cluster singles and doubles) and particularly CCSD(T) (like CCSD but with some triple excitations included), are considered the `gold standard' of computational chemistry; they have resource requirements scaling as $\mathcal{O}(N^6)$ or $\mathcal{O}(N^7)$ respectively, where $N$ is the number of orbitals in the chosen problem basis set.
Very large basis sets are required to obtain energies converging to the complete basis set limit and to reach agreement with experimental values.
This limits the applicability of current-day computational methods to relatively small chemical molecules, and accurate simulations of more complex chemical problems, like catalysed reactions, or large systems like polymers or proteins, remain out of reach. 

The very high computational complexity of simulating chemistry on classical computers has spurred a large interest in simulating quantum chemistry on quantum processors.
The idea of using quantum systems, like quantum computers, to describe quantum systems, like chemical molecules, dates back to the 1980s \cite{Feynmann1982}.
Indeed comparably few, on the order of $10^2$, qubits could be sufficient to demonstrate a practical quantum advantage in this area while the largest commercially available gate-based quantum computer today already has 53 qubits \cite{ibm53}.
We refer to Ref.~\cite{guzik} for an extensive overview on the application of quantum computers to chemistry.
In short, state-of-the-art algorithms for simulating chemistry on a quantum computer can be roughly categorized into near-term NISQ (Noisy Intermediate Scale Quantum \cite{Preskill2018quantumcomputingin}) compatible and future Fault Tolerant Quantum Computing (FTQC) algorithms.
Simulating ground state energies on a FTQC can be done with a variety of algorithms, where (several variants of) the Quantum Phase Estimation (QPE \cite{kitaev1995quantum, Aspuru-Guzik1704}) algorithms are a promising candidate.
QPE may in principle simulate spectra and dynamics of chemistry Hamiltonians to high accuracy; however, the coherence requirements are much more stringent than present-day quantum devices allow for. 

Conversely, in the NISQ era, the so far most widely used algorithm for ground-state estimation is the Unitary Coupled Cluster with Single and Double excitations Variational Quantum Eigensolver (UCCSD-VQE) \cite{VQE}.
UCCSD-VQE is a variational technique in which a state approximate to the ground state is prepared first, such as a Hartree-Fock state, after which a suitable ansatz of parametrized gates is applied to it in a variational hybrid quantum-classical approach in order to prepare a better approximation of the true ground state by tuning the parameters of the ansatz.
The UCCSD method can be seen as a unitary analog to the classical-computational chemistry CCSD protocol \cite{helgaker}, and additionally due to its variational nature is expected to give better accuracy within the same basis set.
Variational CCSD protocols exist on classical computers, but they are formally exponentially-scaling and approximations are required to make them polynomially costly \cite{kutzelnigg}.
Classical computers can perform Unitary CC methods too, but again this is exponentially costly. 

Using a Gaussian orbital basis decomposition of the wavefunction with a Hamiltonian expressed in second-quantization, the quantum circuit depth scaling of the UCCSD protocol is $\mathcal{O}(N^4)$ in state-of-the-art algorithm proposals \cite{McClean_2016}.
Further, the number of measurements has a pre-factor scaling between $\mathcal{O}(N^3)$ and $\mathcal{O}(N^4)$ due to the large number of non-commuting Hamiltonian terms.
Using a particular dual to a plane-wave basis decomposition for the Hamiltonian \cite{babbush} in combination with a fermionic swap-network \cite{swapnetwork} allows for implementing Trotterized operator evolution on a linear array of qubits with depth scaling $\mathcal{O}(N)$, and a number of required terms to perform tomography scaling as $\mathcal{O}(N^2)$.
However, a periodic basis set may be ill-suited for simulating molecular chemistry; the construction of pseudo-potentials analogous to those in conventional plane-wave decompositions is not proposed yet; and a chemistry-inspired quantum ansatz in a periodic basis set has of yet not been identified. 

Given the constraints of NISQ devices it is highly desirable to further reduce the quantum resource requirements of UCC-like variational algorithms.
Taking the ideas developed in classical computational chemistry seriously and translating them to the quantum computing world can be a good guiding principle.
In this paper, we present an efficient quantum algorithm for preparing eigenstate wave functions and energies of molecular systems, using the well known paired-electron approximation, in which electrons are considered as singlet pairs rather than individual fermions. The number of single-occupied spatial orbitals is called the seniority of that determinant. Therefore, if we consider all spatial orbitals are either empty or doubly occupied, this would constitute a seniority-zero Hilbert subspace, whose exact solution is given by the Doubly Occupied Configuration Interaction (DOCI) method \cite{doci1,doci2,doci3,doci4}.
In this approximate mapping, the simulation can be executed with a two-fold increase in simulable system size or basis set, using the same quantum hardware resources, and much fewer total qubit measurements.

Here the implementation of this mapping on a quantum computer is combined with a pair-excitation unitary coupled cluster ansatz with an efficient Trotter step circuit decomposition, in order to simulate molecular chemistry with a circuit depth reduction to $\mathcal{O}(N)$ from $\mathcal{O}(N^4)$ when compared to regular unitary coupled cluster techniques.
We call this ansatz paired-electrons Unitary Coupled Cluster with Double excitations (pUCCD), in reference to prior work using this type of UCC ansatz approach based on fermionic excitation operators (see e.g. Refs.~\cite{kUpCCGSD, Sokolov2020}).
pUCCD can be seen as a unitary and variational quantum version of the classical pair coupled cluster doubles (pCCD) ansatz (also known as Antisymmetric Product of 1-reference orbital Geminal, or AP1roG) which gained recent attention in the field of classical computational chemistry \cite{pccd1,pccd2,pccd3,pccd4,pccd5,pccd6,pccd7,pccd8,pccd9}.

A related (quantum-computational) work \cite{kUpCCGSD} recently also proposed a linear-depth method using a paired-electron approximation.
Although that work naturally enables systematic improvements through the parameter $k$, the paired-electron restriction was not applied to the Hamiltonian as we do in this work, which means a factor of two difference in qubit number requirements of the respective mappings (which has implications for the basis set size or number of included orbitals).
In addition, the Hamiltonian measurement of the energy is vastly simplified in the method presented here.

In this work we concentrate on variational algorithms for near-term quantum computers, although the mapping and ansatz are also well-suited for future FTQC devices.
The reduction in accuracy, as compared to the full Hilbert space can be observed as a trade-off between the gate depth and the qubit number requirements.
We also demonstrate that the paired-electron approximation allows for increased accuracy given a fixed set of available qubits, while simulating the ground state of the lithium hydride molecule. 
Finally, we show how the proposed variational method also presents a faster convergence during optimization of the VQE by several orders of magnitude with respect to the total number of quantum circuit calls than conventional UCCSD-VQE.

\section{Methods}
\label{sec:methods}
\subsection{Hamiltonians in computational chemistry}
To turn a chemical problem into a well defined computational problem, usually a series of assumptions and approximations are made. The first standard assumption is the Born-Oppenheimer approximation, which decouples the nuclear and electronic degrees of freedom.
Considering that electrons move much faster than nuclei in molecules, the Born-Oppenheimer approximation considers the latter as frozen point charges in space and their coordinates hence as classical parameters.
At this point, an electronic structure problem comes up. This corresponds to a problem of determining the behavior of the electrons of the system in the potential created by the nuclei and the background charges.
For an introduction to the electronic structure problem, and a description of classical and quantum algorithms which solves these problems, we refer the reader to Refs.~\cite{mcardle,guzik}.
All wave function based methods require a discretization of the electronic structure problem, which a priori is a continuum problem of electrons in three dimensional space. 

In the standard approach the electrons of the system are modeled individually and finite set of basis functions is introduced (a so-called basis set) to discretize the problem and thereby make it amenable to a solution on a digital computer.
Each basis function is considered to represent a fermionic mode.

The resulting many-body Hamiltonian, acting on the Fock space over these modes, can then be written in second quantized form as follows \cite{guzik}
\begin{equation}\label{fermionicHamil}
\hat{H} = C + \sum_{p,q} h_{p,q} \hat{a}_p^\dagger\hat{a}_q+\frac{1}{2}\sum_{p,q,r,s} h_{p,q,r,s} \hat{a}_p^\dagger\hat{a}_q^\dagger\hat{a}_r \hat{a}_s  ,
\end{equation}
where $C$ is a constant energy offset, $\{p,q,r,s\}$ are fermionic mode indices, $\hat{a}_p$ is the fermionic annihilation operator of the $p$-th fermionic mode, and $h_{p,q}$ and $h_{p,q,r,s}$ are Hamiltonian matrix elements which are determined by means of integrals involving the basis functions and potentials of the nuclei, for example using Hartree-Fock theory or self-consistent-field methods \cite{pyscf}.
The number of terms in this Hamiltonian is $\mathcal{O}(N^4)$, where $N$ is the number of orbitals.

The number of fermionic modes and the values of the matrix elements depend on the chosen basis set.
In a minimal basis set, like the STO-6G basis, as many basis functions are used for core and for valence orbitals of the molecule under study.
In the case of the lithium hydride (LiH) molecule, which we consider as an example in this paper, the number of basis functions is $6$ using this basis. 
Electrons being spin-1/2 particles, two of them can occupy a single fermionic mode (spin-up and spin-down).
This brings the total number of spin-orbitals (SOs) to $12$.

The fermionic Hamiltonian in \eqnref{fermionicHamil} can be mapped to qubits on a gate-based quantum computer with the help of a variety of transformations (such as the Jordan-Wigner transformation \cite{Jordan1928}).
The result of this is a Hamiltonian consisting of a sum of Pauli string operators, i.e., tensor products of the Pauli operators $\hat{\sigma}^\text{X}$, $\hat{\sigma}^\text{Y}$, and $\hat{\sigma}^\text{Z}$ and the $2 \times 2$ identity operator.

\subsection{Discretization in the paired-electron approximation}
In this work, we employ the paired-electron approximation \cite{helgaker} during the discretization step.
This approximation effectively restricts the problem to a zero-seniority subspace of the full Hilbert space of all possible many-body electronic states that is spanned by those states in which all orbitals are either empty or occupied by an electron singlet pair (because of the Pauli exclusion principle).
The paired-electron assumption is rather widespread in computational chemistry and has proved to deliver sufficient accuracy for the description of a range of chemical systems.
Indeed its usage has become standard in computational chemistry \cite{jensen}.
This is so, as quite often in molecular systems all the electrons are distributed in electronic pairs occupying the same spatial orbital.  

In this approximation the Hamiltonian \eqnref{fermionicHamil} can be written as
\begin{equation}\label{bosonicHamil}
\hat{H}_r = C + \sum_{p,q} h^{(r1)}_{p,q} \hat{b}_p^\dagger\hat{b}_q + \sum_{p\neq q} h^{(r2)}_{p,q} \hat{b}_p^\dagger\hat{b}_p \hat{b}_q^\dagger\hat{b}_q,
\end{equation}
where $\hat{b}_p$ represents the electron-pair annihilation operator in mode $p$, where each mode is limited to either $0$ or $1$ excitation.
The matrix elements $h^{(r1)}_{p,q}$ and $h^{(r2)}_{p,q}$ are related to the one- and two-electron integrals (see the Appendix for details and f.ex. Ref.~\cite{pccd8} for a derivation).
We in this work focus on closed-shell systems, with total spin zero, but extensions can be made which introduce Heisenberg-like spin terms in \eqnref{bosonicHamil} \cite{pccd8}.
The `electron-pair annihilation operator' $\hat{b}_p$ can be defined through the hard-core boson (HCB) \mbox{(anti-)}commutation relations \cite{CIP361}
\begin{eqnarray}
[\hat{b}_p, \hat{b}_q^\dagger] = [\hat{b}_p^\dagger, \hat{b}_q^\dagger] = [\hat{b}_p, \hat{b}_q] &=& 0 \hspace{15pt} (p\neq q)\\
\{\hat{b}_p\dagger, \hat{b}_p^\dagger\} = \{\hat{b}_p, \hat{b}_p\} &=& 0\\
\{\hat{b}_p, \hat{b}_p^\dagger\} &=& 1
\end{eqnarray}
The connection here is that a singlet pair of electrons has even spin and thus behaves like a boson, while no more than 2 electrons can occupy any of the molecular orbitals at once.
Notice that this Hamiltonian is far simpler than the unrestricted fermionic Hamiltonian in \eqnref{fermionicHamil}, but still non-trivial, as it is not quadratic in the creation and annihiliation operators. 

As a first consequence of using the paired-electron approximation, the number of qubits needed to represent a given problem is halved compared with the unrestricted mapping.
In the LiH example the $6$ basis functions in the STO-6G basis map to $6$ qubits instead of the usual $12$.
Alternatively, one may use the same qubit number with a larger basis set (this can be advantageous as we will show later).
For example, in the 4-31G basis set, $11$ molecular orbitals (MOs) result in $22$ fermionic orbitals but only $11$ qubits are needed in the paired-electron approach.
In this way, with $12$ qubits as a quantum resource, one may either simulate LiH in unrestricted STO-6G or in paired-electron 4-31G. 

As the electron pair creation and annihilation operators commute, the above Hamiltonian can be mapped directly to Pauli spin operators resulting in operations native on qubit based gate quantum computer using the representation
\begin{equation}\label{transformationrules}
\hat{b}_p = \frac{1}{2}(\hat{\sigma}^\text{X}_{p}+i\hat{\sigma}^\text{Y}_{p}),
\end{equation}
where $\hat{\sigma}^\text{X}_{p}$ and $\hat{\sigma}^\text{Y}_{p}$ are Pauli-X and Y spin operators respectively, showing a well-known correspondence between HCB operators and Pauli-spin operators \cite{bratteli}.

Due to the hermiticity of \eqnref{bosonicHamil}, some terms vanish, and the total qubit Hamiltonian can be written as

\begin{figure*}
	\centering
	\includegraphics[width=0.9\linewidth]{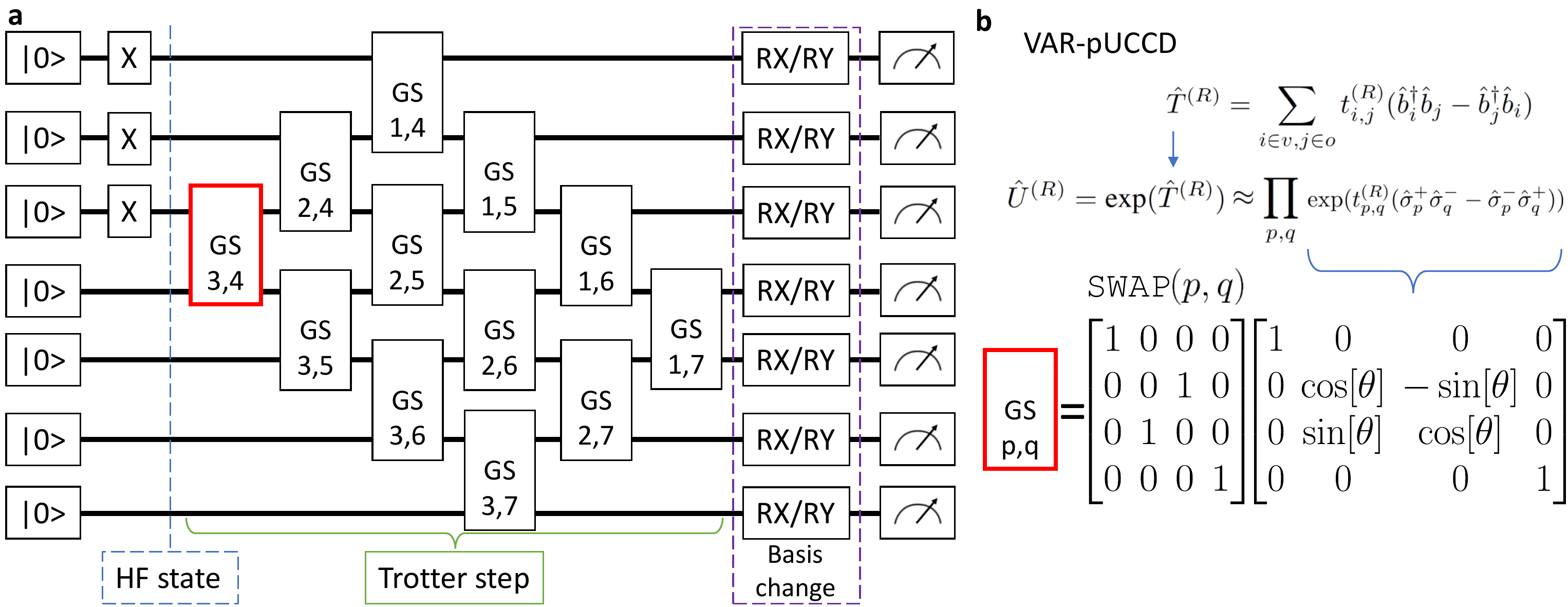}
	\caption{\textbf{a} Quantum circuit diagram implementing a single Trotter step of the pUCCD ansatz, assuming a nearest-neighbour coupled qubit architecture. The qubit register is first prepared in the HF state by the application of X gates to the first $n_e$ qubits. The pUCCD ansatz Trotter step is then implemented as a sequence of Givens-SWAP (GS) gates in linear circuit depth. In a VQE implementation, the energy is estimated by Hamiltonian averaging, which is performed by rotating all the qubits to the relevant basis for those Hamiltonian Pauli-terms and measuring the qubits in their Z-basis.  \textbf{b} The GS gate is shown as a mathematical representation, and the origins of the pUCCD coupled cluster operators is detailed. The GS gate is a combination of a Givens rotation, or parametrized swap gate, followed by a full swap gate. In hardware architectures with all-to-all connectivity, the swap-network as shown can be replaced by only the parallel Givens rotations.}\label{circuit_diagram}
\end{figure*}

\begin{align}\label{qubithamil} 
\hat{H}_{qb} =C &+ \sum_{p}  \frac{h^{(r1)}_{p}}{2} (\hat{I}_{p}-\hat{\sigma}^\text{Z}_{p})\\
&+ \sum_{p\neq q} \frac{h^{(r1)}_{p,q}}{4}
(\hat{\sigma}^\text{X}_{p}\hat{\sigma}^\text{X}_{q}+\hat{\sigma}^\text{Y}_{p}\hat{\sigma}^\text{Y}_{q}) \nonumber\\
&+ \sum_{p\neq q} \frac{h^{(r2)}_{p,q}}{4}   (\hat{I}_{p}-\hat{\sigma}^\text{Z}_{p}-\hat{\sigma}^\text{Z}_{q}+\hat{\sigma}^\text{Z}_{p}\hat{\sigma}^\text{Z}_{q})\nonumber
\end{align}
where $\hat{\sigma}^\text{Z}_{p}$ and $\hat{I}_{p}$ are Pauli-Z and identity spin operators respectively. \eqnref{qubithamil} belongs to the class of HCB Hamiltonians, for which approximating the ground state energy is known to be QMA complete \cite{Childs2014}.
QMA is the complexity class Quantum Merlin Arthur, and this essentially means that it is extremely unlikely that ground states energies of general Hamiltonians of this form (\eqnref{qubithamil}) can be found efficiently.
It is thus also unlikely (although not entirely impossible) that the Hamiltonians arising from chemistry problems have a special structure that allows highly-accurate approximations to the ground state to be computed efficiently.

As is apparent from both \eqnref{bosonicHamil} and \eqnref{qubithamil}, the total number of local terms in the paired-electron Hamiltonian scales as $\mathcal{O}(N^2)$ (as opposed to $\mathcal{O}(N^4)$ in the unrestricted case).
This, together with the fact that the Hamiltonian in \eqnref{qubithamil} naturally decomposes into just two non-commuting components, greatly simplifies the measurement of the energy in VQE schemes based on Hamiltonian averaging. 
In fact, the Hamiltonian terms can be grouped into just three unique tensor product bases. Effective measurements in these bases can be achieved using just single-qubit rotations.

\subsection{Trial state}
A first step of both VQE and QPE algorithms, and most Hamiltonian simulation algorithms in general, is the preparation of a trial state.
The success of an algorithm for determining eigenenergies of the Hamiltonian depends on the quality of the state preparation and its closeness to the actual eigenstate of interest.
A good initial trial state for the ground state of $\hat{H}_r$ is the Hartree-Fock state, which in this case is just a product state with the $n_e$ lowest-energy MOs occupied with a single pair of electrons.
For systems with an even number of electrons, the HF state in the paired-electron approximation represents the same state as without the approximation, and the energy expectation is equal too. 

Such a simple trial practically never captures the complexity of the true, partially entangled ground state.
Therefore, after preparing the HF-state on a qubit lattice, an additional (variational) circuit can be applied to the qubits which prepares a more general ansatz including higher-order correlators.
In principle, an eigenstate of the many-body Hamiltonian could contain many complicated correlations, which implies that a large number of entangling operations need to be applied to the HF state (which is by itself a mere product state) in order to produce this highly entangled state.
In practice however, single/double/triple single-particle excitations are often enough to bring the HF state sufficiently close to the ground state.
This idea is behind the coupled-cluster method, where typically single, double and triple excitations are considered, like in the CCSDT and CCSD(T) ans\"atze.

In classical computational chemistry, a non-unitary Coupled Cluster operator is applied to the HF state as an approximation, because computing the exponential of a unitary matrix can be very costly.
However, on a quantum computer the unitary evolution over a coupled-cluster operator can be naturally implemented.
In conventional UCCSD methods the unitary operator is constructed as follows \cite{UCCSD}:
\begin{align}
\hat{U}&=\exp (\hat{T}_1+\hat{T}_{2,1}+\hat{T}_{2,2}) \label{uccsd_operator}\\
\hat{T}_1&=\sum_{i\in v,j\in o}t^{(1)}_{i,j} (\hat{a}^\dagger_i \hat{a}_j -\hat{a}_j^\dagger \hat{a}_i)\\
\hat{T}_{2,1}&=\sum_{i\in v,j\in o}t_{i,j}^{(2,1)} (\hat{a}^\dagger_i  \hat{a}_j \hat{a}^\dagger_{i+1}  \hat{a}_{j+1} -\hat{a}^\dagger_j  \hat{a}_i \hat{a}^\dagger_{j+1}  \hat{a}_{i+1})\\
\hat{T}_{2,2}&=\sum_{(p,q)\in v,(r,s)\in o}t_{p,q,r,s}^{(2,2)}( \hat{a}^\dagger_p  \hat{a}_q \hat{a}^\dagger_{r}  \hat{a}_{s} -\hat{a}^\dagger_q  \hat{a}_p \hat{a}^\dagger_{s}  \hat{a}_{r})
\end{align}
where $\{i,j\}$ and $\{p,q,r,s\}$ are single- and two-particle fermionic mode indices respectively, $\{v,o\}$ are the sets of virtual and occupied orbitals respectively, $\hat{a}_p$ is the fermionic annihilation operator of the $p$'th fermionic mode, $t^{(1)}_{i,j}$ refers to excitations involving two orbitals and a single electron,  $t^{(2,1)}_{i,j}$ refers to excitations involving two distinct orbitals and two electrons, and $t^{(2,2)}_{p,q,r,s}$ refers to excitations involving up to four distinct orbitals and two electrons.

To simulate the unitary evolution operator \eqnref{uccsd_operator} with a quantum circuit based on single and two-qubit gates, one can Trotterize \cite{trottersuzuki} the evolution and apply each term sequentially or in a structured \cite{trotterCC, trotterCC2} or random \cite{trotterCC3,trotterCC4} way.
If the amplitudes are not too big, this Trotterization leads only to small errors compared to the true UCCSD state.
Often a single Trotter step can be sufficient to accurately and efficiently reproduce the ground-state energy of simple molecular systems \cite{oneTrotterOkay} and the robustness inherent to VQE may even partially compensate the Trotter errors \cite{vqeRobust}.

Within the paired-electron approximation, the UCC ansatz only includes gates that move pairs of electrons between MOs.
We may write such an ansatz in the paired-electron `bosonic' operator picture:
\begin{align}
\hat{U}^{(R)}&=\text{exp}(\hat{T}^{(R)})\label{exactunitary}\\
\hat{T}^{(R)}&=\sum_{i\in v,j\in o}t^{(R)}_{i,j}( \hat{b}^\dagger_i \hat{b}_j -\hat{b}_j^\dagger \hat{b}_i) \\
\hat{T}^{(Q)}&=\sum_{p,q} t_{p,q}^{(R)}(\hat{\sigma}^{+}_p  \hat{\sigma}^{-}_q -  \hat{\sigma}^{-}_p  \hat{\sigma}^{+}_q)\\
&=\sum_{p,q} \frac{t_{p,q}^{(R)}}{2} (\hat{\sigma}^{X}_p  \hat{\sigma}^{Y}_q -  \hat{\sigma}^{Y}_p  \hat{\sigma}^{X}_q) \label{RUCCqubitUnitary}
\end{align}
where $\hat{U}^{(R)}$ represents the coupled cluster unitary restricted to only include paired-electron evolutions, $\hat{T}^{(R)}$ is the paired-electron coupled cluster operator, the superscript $(R)$ denotes the electron-singlet restricted (paired-electron) approximation, and we transform  $\hat{T}^{(R)}$ to a qubit-acting cluster operator $\hat{T}^{(Q)}$ using the transformation rule of \eqnref{transformationrules}.
The form of this ansatz has been considered previously in analogous studies, which maintain the electronic fermion-operator structure \cite{kUpCCGSD, Sokolov2020}, where this ansatz is also referred to as pUCCD ansatz. 
From here on we refer to the pUCCD as the ansatz constructed from hard-core boson operators representing paired-electron excitations. 
From the form of the ansatz, it is clear that the unitary operation is particle and spin conserving.
There are $(N-n_e)n_e$ terms, where $n_e$ is the number of electron pairs in the system and $N$ is the number of orbitals.

A first-order Trotterization of evolution over the qubit operator \eqnref{RUCCqubitUnitary} is given by
\begin{align}
\hat{U}^{(R)}=\exp(\hat{T}^{(Q)}) &= \text{exp}(\sum_{p,q} t_{p,q}^{(R)}(\hat{\sigma}^{+}_p  \hat{\sigma}^{-}_q -  \hat{\sigma}^{-}_p  \hat{\sigma}^{+}_q))\\
&\approx\prod_{p,q} \exp (t_{p,q}^{(R)}(\hat{\sigma}^{+}_p  \hat{\sigma}^{-}_q -  \hat{\sigma}^{-}_p  \hat{\sigma}^{+}_q)) \label{approxTrotter}
\end{align} 
where each $\exp (t_{p,q}^{(R)}(\hat{\sigma}^{+}_p  \hat{\sigma}^{-}_q -  \hat{\sigma}^{-}_p  \hat{\sigma}^{+}_q))$ can be implemented as a partial swap gate between qubits $p$ and $q$. The order of applications of the product of unitaries in \eqnref{approxTrotter} matters and can be chosen optimally to minimize the Trotter error \cite{trotterCC, trotterCC2}.

In \figref{circuit_diagram} we show how the above unitary can be simulated in linear circuit depth on a quantum computer with a linear array of qubits using a discrete set of pre-programmed unitary operations (or ``gates''), which are subsequently variationally optimized to prepare an approximation to the ground state.
The qubits are initialized in the $|0\rangle$ state and the lowest $n_e$ orbitals are populated via an $X$-gate.
The pUCCD circuit is then executed.
In \figref{circuit_diagram}\textbf{a} we show the corresponding quantum circuit diagram for a single Trotter step (this can readily be extended to larger numbers of Trotter steps).
In \figref{circuit_diagram}\textbf{b} we detail the Givens-SWAP (GS) operation (``gate'').
The GS-gates are Givens rotation (parametrized-swap) gates followed by a full-swap gate. The full-swap gate $\texttt{SWAP}(p,q)$ swaps the qubit labels in order to bring every qubit which was occupied next to every other qubit which was not occupied originally.
In this way, excitations from every occupied orbital to every virtual orbital is simulated, in a minimal gate-depth of $N$, even on a linear chain of qubits (nearest-neighbor connectivity) \cite{swapnetwork}.

We note here that Ref.~\cite{Nam2020} also partially employs the concept of treating pairs of electrons into hard-core bosonic modes, reducing the circuit complexity of implementation and initially using half of the qubits.
However, in this work the Hamiltonian is not modified and instead one returns to a spin orbital mapping using CNOT gates after having performed the double-excitation circuit.
This means that the Hamiltonian in this work still has the Jordan-Wigner term grouping issues, incurring a significantly larger overall runtime.

The pUCCD ansatz shows resemblances with classical pCCD. In recent years \cite{pccd1,pccd2,pccd3,pccd4,pccd5,pccd6,pccd7,pccd8,pccd9}, this method has become a promising direction in order to reach near-DOCI accuracies, as the latter can be seen as the exact ground state within the seniority-zero Hilbert space but with a factorially-scaling runtime.
DOCI is able to properly account for non-dynamic or static correlations, dominant in bond breaking and forming processes, for instance.
Although zero-seniority Hilbert subspaces fail at recovering dynamic correlation, they provide very useful insights on how to efficiently treat correlations in higher-seniorities subspaces\cite{doci5}.
pCCD, being spanned in a zero-seniority subspace, has DOCI as a limit.
Concomitantly, it has been shown that pCCD does not fully recover DOCI, which can be compensated by including extra terms in the coupled-cluster expansion \cite{pccd8,pccd9}.
Finally, conventional pCCD is non-unitary and non-variational.
As previously shown, pUCCD reproduces quite closely DOCI, even for situations where non-dynamic correlations are significant.

Now we briefly comment on the complexity of simulating the hard-core bosonic pUCCD ansatz quantum circuit on classical computers. 
It is known that the output of certain circuits acting on product states, consisting of only nearest-neighbour Givens rotations, can be computed efficiently on a classical computer \cite{matchgates2008}. 
This is also the case with the Hartree-Fock ansatz circuit used in Ref.~\cite{arute2020hartreefock}, which has a very similar structure to the pUCCD ansatz.
However, this efficient-computability vanishes when considering non-nearest neighbour Given rotations, which we effectively perform via the SWAP network. 
In hardware with all-to-all connectivty, one would not need the SWAP network and could perform the all-to-all Givens rotations directly. 
Still in that case such a circuit could not be simulated efficiently classically with the methods from Ref.~\cite{matchgates2008}. 

\subsection{Measuring the energy}
After performing the state preparation with a parametrized circuit (with parameters $t_{p,q}^{(R)}$), one may calculate the energy of the prepared state in different ways:
one possibility is to use one of the Quantum Phase Estimation methods \cite{kitaev1995quantum, nielsen_chuang_2010} (which yield arbitrarily good precision but have too stringent coherence requirements for current NISQ hardware), or to use Hamiltonian averaging.
We here focus on the latter, an implementation of a Variational Quantum Eigensolver (VQE \cite{VQE}), although the pUCCD ansatz may in general be applied in any simulation strategy involving quantum chemistry state preparation.

In the measurement phase, one can only measure sets of mutually commuting operators simultaneosly due to the fundamental constraints imposed on measurement by the laws of quantum mechanics.
In conventional UCCSD-VQE, the number of such sets scales as $\mathcal{O}(N^4)$, increasing the overall computation time, which may be brought down to an upper bound of $\mathcal{O}(N^6)$ measurements for most realistic molecules in a Gaussian type orbital basis set (see Ref.~\cite{measurements}).
However, in the paired-electron method described in this paper, all Hamiltonian operators can be sorted into just $2$ groups of mutually commuting sets.
If we assume that only single-qubit rotations are used for diagonalizing the Pauli strings, there are just $3$ unique tensor product bases (I, Z and ZZ terms, XX terms, and YY terms).
This gives only a constant overhead, $\mathcal{O}(1)$, in the overall complexity due to the measurement phase contribution, with a worst-case-scenario scaling upper bound of $\mathcal{O}(N^4)$ measurements due the repetitions required to reach a fixed desired accuracy despite shot noise.
The final total time scaling of pUCCD is then estimated to have a polynomial speedup over state-of-the-art quantum simulation protocols.

\section{Results and Discussion}
We now illustrate the applicability and advantages of the paired-electron approximation.
We test our simulation method by applying it to the ground state energy estimation of the lithium hydride molecule.
The lithium hydride molecule has been studied extensively using exact methods in conventional computational chemistry, because the ground state structure is at least a little bit more complicated at all bond lengths than that of very simple molecules such as dihydrogen and helium-hydride.
As such, the Hartree-Fock state has insufficient overlap with the ground state and a good ansatz operator is required to reach chemical accuracy with the exact ground state energy within a given basis.
It therefore presents a non-trivial, yet simple and exemplary test case for near-term quantum computational chemistry algorithms.

\begin{figure}
	\centering
	\begin{tabular}{ l   }
		\textbf{a}\\
		\includegraphics[width=0.97\linewidth]{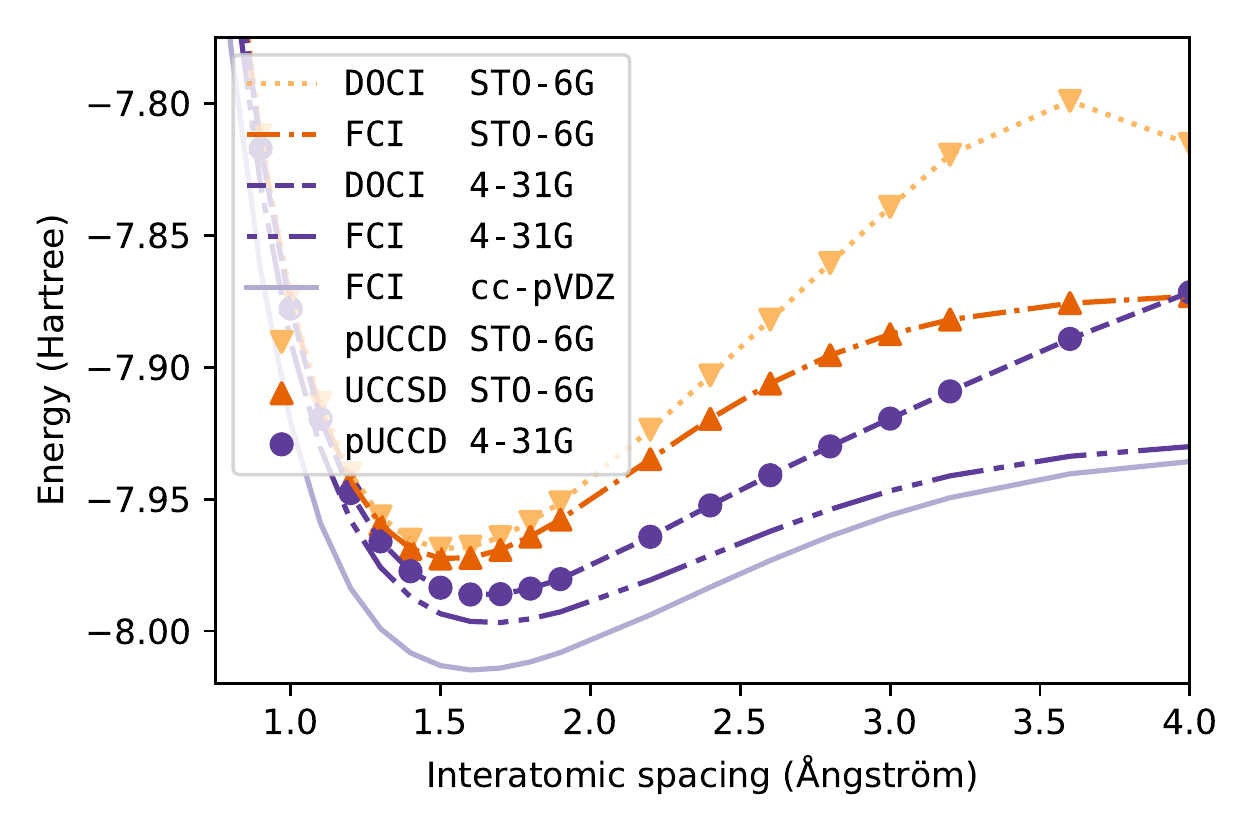} \\
		\textbf{b}\\ 
		\includegraphics[width=0.95\linewidth]{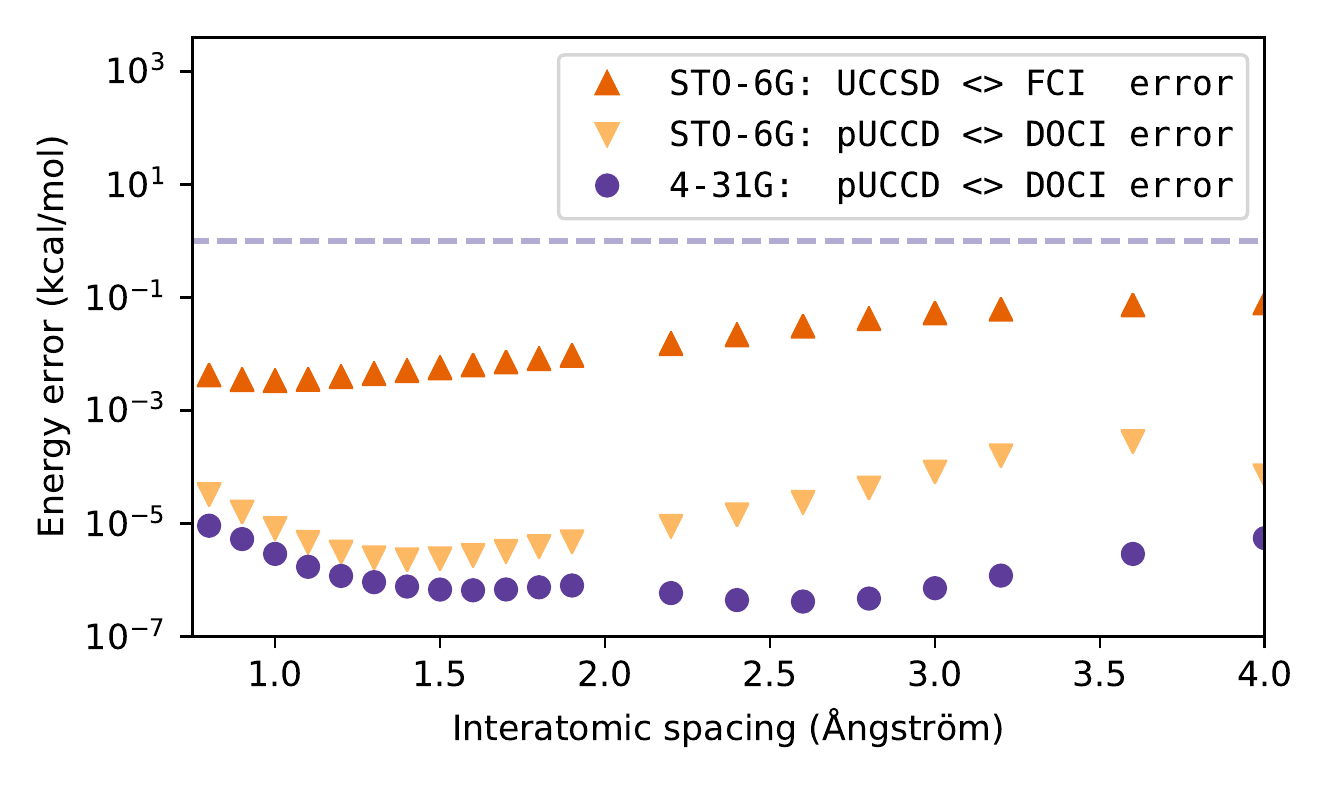}  
	\end{tabular}
	\caption{\textbf{a} Ground state energy of the lithium hydride molecule as a function of interatomic distance. The light-purple (solid) curve depicts a highly accurate cc-pVDZ basis set FCI solution, the dark-orange (dash-dotted) curve depicts the FCI solution in STO-6G basis, the yellow (dotted) curve depicts the DOCI solution in STO-6G basis, and the dark-purple (dashed) curve depicts the 4-31G basis DOCI solution. Dark-purple circles, dark-orange triangles and yellow triangles depict the pUCCD-VQE in 4-31G, UCCSD-VQE in STO-6G and pUCCD-VQE in STO-6G converged optimizer results, respectively. \textbf{b} Difference between converged pUCCD and DOCI energies in the 4-31G and STO-6G basis sets, and difference between UCCSD and FCI energies in the STO-6G basis set, as a function of interatomic distance. Chemical accuracy, 1 kcal/mol, is plotted for reference with a light-purple (dashed) line.}\label{restricted_plot}
\end{figure}

\subsection{Comparison of converged results without noise}
We start with simulation results obtained via a noise free evolution on a quantum computer simulator.
Firstly, we will focus on the expressive power of the pUCCD ansatz and will concentrate on issues of convergence and learning of the parameters later.
Therefore, we initialize the angles based on the CCSD amplitudes found with PySCF \cite{pyscf} and compute the energy by calculating the exact Hamiltonian expectation value based on the simulated circuit wavefunction.
In the simulation we have access to the full wave function and therefore the resulting expectation value of the energy is deterministic.
We thus here do not consider the issue of how to optimally decompose the Hamiltonian into operators that can be measured on a real (non-simulated) quantum processor \cite{huggins2019}.
It should be noted however, that due to the lower number of non-commuting terms, also here the pUCCD method has advantages over the standard UCC method. 

We then optimize the parameters of the ansatz circuit, shown in \figref{circuit_diagram}, and thereby find the pUCCD energy, which is an upper bound to the true ground state energy (in the respective basis and level of approximation).
We compare the obtained energy with the exact ground state energy obtained by FCI, allowing for any combination of single- or double electron excitations, and the ground state energy of the Hamiltonian in the zero-seniority subspace obtained by DOCI.

In \figref{restricted_plot} we show the simulation results for the lithium hydride molecule at various interatomic distances.
We compare the UCCSD and pUCCD methods with FCI and DOCI, respectively.
We use two different basis sets, namely STO-6G (6 MOs) and 4-31G (11 MOs).
As additional reference FCI results in the large cc-pVDZ basis set (19 MOs) are also included.
For the pairing scheme, we chose the intuitively straightforward restricted Hartree-Fock (RHF) method, although broken-symmetry geminals can in principle also be used and previous work has shown that this can generally be much improved, yielding higher accuracies for paired-CC approaches \cite{pccd5, pccd6,JOHNSON2013101}.
Simulations of the UCCSD and pUCCD VQE yield very good agreement with the corresponding exact ground state energy (FCI and DOCI) in the same basis set, over the whole range of interatomic distances.

When comparing our results to FCI results obtained from an unrestricted diagonalization in the much larger cc-pVDZ set, we find that the exact ground state energy in paired-electron approximation with the 4-31G basis is much closer to cc-pVDZ, than that in the unrestricted case with the STO-6G basis.
The error introduced by the electron singlet approximation is much smaller than the accuracy gain due to the larger basis set.
This is particularly noteworthy as the unrestricted STO-6G computation requires a comparable (and even slightly larger) number of qubits to the paired-electron 4-31G calculation ($12$ versus $11$ qubits, respectively).
We note here that strictly speaking one should not directly compare energies between different basis sets in a practical quantitative analysis. We here merely illustrate the potential for an advantage in releasing additional quantum resources and using them for an increased number of basis functions, which we estimate will typically allow a better ultimate limit to the accuracy set by the basis set.
The additional error from using a single Trotter step VQE instead of FCI is well over a thousand times less than chemical accuracy (see \figref{restricted_plot} bottom), which is the precision achieved typically in chemical experiments.
All in all, we find that the pUCCD accuracy in the larger basis set is higher than the UCC accuracy in the smaller basis set, while at the same time requiring fewer quantum resources ($11$ instead of $12$).
That shows that pUCCD is able to make better use of limited quantum resources in order to obtain more accurate results than UCCSD-VQE.
We note that we have not applied any qubit reduction schemes, for example based on symmetries in the problem \cite{bravyi2017tapering} or exploiting block-diagonality \cite{Moll_2016}.
Those techniques can in principle also be used in the mapping used here, allowing a further decrease of computational resources.

\subsection{Convergence and training}
We now compare the convergence of the pUCCD-VQE optimization experiment with that of conventional UCCSD-VQE.
So far we considered exact expectation values, but in practice one should consider the stochastic nature of Hamiltonian averaging process. 
The required number of shots is inversely proportional to the desired accuracy, squared.
For the optimization of the VQE experiments, we set the number of shots such that the standard deviation was half chemical accuracy, i.e., $\frac{1}{2}\times0.0016$ Hartree which is $0.5$ kcal/mol.
The statistics were estimated by means of a pre-calculation with the ansatz initialized at the CCSD angles setting for both the pUCCD and the UCCSD experiments, resulting in an estimated need of $41,000$ and $100,000$ shots respectively.
We used the Implicit Filtering algorithm \cite{ImFil} for the classical optimization to cope with the derivative-free, noisy blackbox function optimization.
\begin{figure}[tb]
	\centering
	%\begin{tabular}{ c   }
	\includegraphics[width=1.0\linewidth]{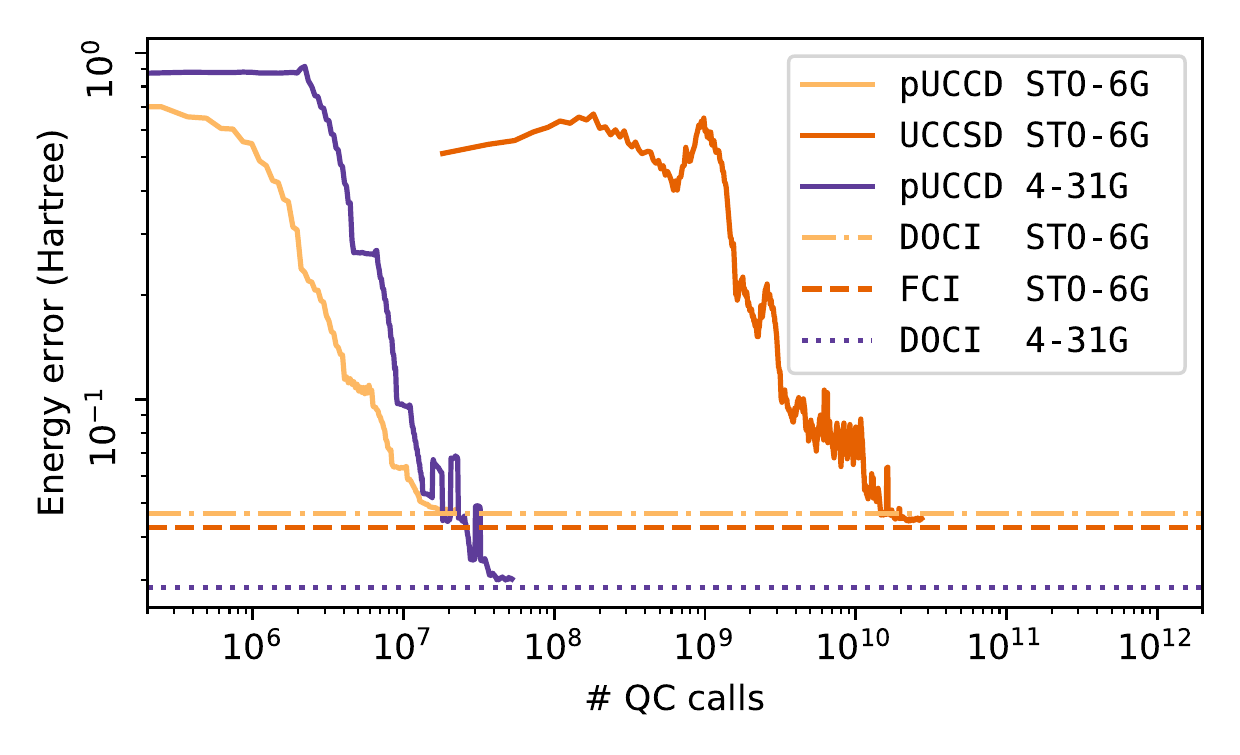}
	%\end{tabular}
	\caption{Convergence of LiH ground state energy simulation using three different VQE Coupled-Cluster implementations. The y-axis shows the energy error (in Hartree) with respect to the FCI ground state energy in cc-pVDZ basis.
		The curves represent the overall convergence behaviour of VQE experiments for UCCSD in STO-6G, pUCCD in STO-6G and pUCCD in 4-31G (dark-orange, yellow and dark-purple curves, respectively).
		The plotted curves are averaged over 10 runs and a 10-point smoothing window is convoluted with it to suppress fluctuations and make the overall behavior more apparent.
		The x-axis expresses the total number of times the quantum circuit is initialized, run and measured.
		As reference lines, we plot the exact FCI and DOCI ground state energies in each respective basis in light-purple (dash-dotted, dashed, and dotted curves) gates.}
	\label{convergence_plot}
\end{figure}

Frist, it can be observed that the pUCCD method requires fewer parameters.
This is because only the $n_{occ}\times n_{virt}$ CC doubles amplitudes are to be considered within the paired-electron approximation.
As in any CC technique, one may do MP2 pre-screening to reduce this number further.
Second, the number of groups of simultaneously measurable Pauli terms are $182$ in the case of conventional UCCSD, while this number is reduced to just $3$ for pUCCD.
Third, we find that pUCCD is very significantly faster to converge.
In \figref{convergence_plot} we plot the results for simulating LiH at equilibrium bond distance, showing the same three basis settings as in \figref{restricted_plot}).
We find that: not only does the pUCCD method in the larger basis set (4-31G) converge to a lower final energy than UCCSD with a similar number of qubits in the smaller basis set (STO-6G), it also does so with much fewer total number of calls to the QPU (shots), in the range of three orders of magnitude.
Also during the training, the pUCCD method consistently gives better results for the same number of shots.
Given the gate and readout times of present and near future quantum computers, such reduction in the total number of quantum experiments can verily be the decisive factor for whether a quantum simulation of a molecule is feasible or not.

Similarly, one may compare the convergence within the same basis set (STO-6G).
Also here the number of calls to the QPU shrinks by nearly three orders of magnitude, but the advantage of converging to a more precise final value is lost, at the advantage of a reduction in qubits needed.

\subsection{Influence of experimental imperfections}

In NISQ-era devices, gate and measurement errors have a significant detrimental impact on the accuracy of computations.
We note that in the above simulations the circuit was executed perfectly without such errors, and the energy estimation was taken over the exact wave function (only shot noise was considered).
Although a full consideration of the noise goes beyond the scope of this work, we may compare pUCCD to conventional UCCSD,  where the pUCCD method has a much smaller total gate count and (two-qubit) circuit depth.
This reduces the influence of decoherence and increases overall fidelity, which means that the pUCCD results should be more robust against noise than general UCC circuits. 
Fermionic pUCCD would incur an additional prefactor in two-qubit gate counts to account for the anti-symmetrization of the electron wavefunctions on-chip.

Many noise mitigation techniques beyond VQE's inherent error suppression have been proposed, such as noise extrapolation techniques, probabilistic error cancellation, quantum subspace expansion, stabilizer based methods and more (for a recent overview, please see e.g. Ref.~\cite{mcardle}).
All these can be used in the present hard-core boson-based pUCCD-VQE method as well.
In particular, the stabilizer method would entail doing checks on the number of electron pairs which is a conserved quantity according to the electronic structure Hamiltonian. Also, all Hamiltonian term measurements which are done in the Z-basis can natively be used to check the electron number without additional circuits. We detail such post-selection technique in the next section.

\subsection{Error-mitigation by post-selection}
When estimating \eqnref{qubithamil} on a wavefunction prepared on the QPU in the presence of noise, we may exploit some of the physical symmetries of the chemistry system we are simulating. In particular, we note the following; the initial reference state, the restricted Hartree-Fock state, has a particular known number of `excitations' (pairs of electrons). Next, the unitary ansatz circuit pUCCD is particle-conserving, which means the number of electron pairs is conserved. The trial state is then some superposition of basis states with that same number of pairs of electrons, resulting in only bitstrings with that many (1's) in the measurement results. However, if errors occur in the circuit or in the measurement phase, this may no longer always be the case. As we know any physical state must conform to the particle-number conservation, we can post-select the measurement results on that condition. This may significantly improve the result. 
However, there are some caveats. For one, the number of particles being correct does not mean \textit{strictly} no errors occurred in that case, just that the number of errors is either zero or they compensate/cancel out to result in the correct pair number. Also, the above strategy only holds true when measuring in the particle-basis, i.e. the Z operators in \eqnref{qubithamil}. 

To tackle the latter caveat, there is a better strategy. We may rotate the qubit array to other bases than X, Y or Z. In particular, we could diagonalize parts of the Hamiltonian acting on pairs of qubits $(i,j)$ as
\begin{equation}\label{rotation_gate}
\mathcal{U}_{p,q}(\pi/4)^\dagger \left( \hat{\sigma}^X_p\hat{\sigma}^X_q+\hat{\sigma}^Y_p\hat{\sigma}^Y_q \right) \mathcal{U}_{p,q}(\pi/4) = \hat{\sigma}^Z_p - \hat{\sigma}^Z_q
\end{equation}
where
\begin{equation}
\mathcal{U}_{p,q}(\theta) = 
\begin{pmatrix} 
1 & 0 & 0 & 0\\ 
0 & \cos \theta & -\sin \theta & 0\\ 
0 & \sin \theta & \cos \theta & 0\\ 
0 & 0 & 0 & 1\\ 
\end{pmatrix}
\end{equation}
is a Givens rotation over angle $\theta$ between qubits $p$ and $q$, with matrix basis ${|00\rangle,|01\rangle,|10\rangle,|11\rangle}$. Or, inversely, we could rotate the system wavefunction before measurement by a circuit-synthesized unitary $\mathcal{U}^\dagger$ and then measure the diagonal $\hat{\sigma}^Z_i$ and $\hat{\sigma}^Z_j$ operators simultaneously. We here show how this helps in measuring the expectation value of the chemistry Hamiltonian in qubit form from \eqnref{qubithamil}. We first divide the task into measuring the expectation value of the XX+YY terms, the Z and ZZ terms, and the constant term $C'$ (which now absorbs the identity terms in addition to the $C$ constant from \eqnref{qubithamil}):
\begin{equation}
E = \langle \psi |\hat{H}_{xx,yy} |\psi\rangle + \langle \psi |\hat{H}_{z,zz} |\psi\rangle + C' 
\end{equation}
where we may efficiently evaluate the term $\langle\hat{H}_{z,zz} \rangle=\langle \psi |\hat{H}_{z,zz} |\psi\rangle$ in one go as the Hamiltonian is diagonal in the qubit operators. We next write the term $\langle \psi |\hat{H}_{xx,yy} |\psi\rangle$ as 
\begin{align}
\langle \hat{H}_{xx,yy} \rangle &= \sum_{p\neq q} \langle \psi|  \frac{h^{(r1)}_{p,q}}{4}(\hat{\sigma}^\text{X}_{p}\hat{\sigma}^\text{X}_{q}+\hat{\sigma}^\text{Y}_{p}\hat{\sigma}^\text{Y}_{q})|\psi\rangle \\
=& \sum_{p\neq q}\langle \psi| \mathcal{U}_{p,q}(\theta)  \frac{h^{(r1)}_{p,q}}{4}( \hat{\sigma}^Z_p - \hat{\sigma}^Z_q ) \mathcal{U}_{p,q}^\dagger(\theta)|\psi\rangle ,
\end{align}
where in the last step, a basis rotation on qubit pairs $\{p,q\}$ is executed that diagonalizes those terms. In effect, this allows to measure the XX and YY terms simultaneously for each pair, \textit{and} to rotate to a diagonal basis where the particle number should have been maintained. If we do this for $N/2$ distinct pairs of the total $N$ modes, we effectively measure all modes in the particle-basis and can therefore filter out some of the noisy bitstring measurements. This is unfortunately not possible for \textit{all} $\mathcal{O}(N^2)$ XX+YY terms simultaneously. This seems to be because the $\hat{b}_p^\dagger\hat{b}_q$ terms in the Hamiltonian are actually two-particle operators in disguise, effectively corresponding to the $\hat{a}_p^\dagger\hat{a}_q^\dagger\hat{a}_r\hat{a}_s$ operators from Ref.\cite{huggins2019} which are diagonalized in $L$ steps, at worst scaling as $\mathcal{O}(N^2)$ but in our case can be efficiently performed in $\mathcal{O}(N)$ by parallelizing the operations assuming all-to-all connectivity in f.ex. ion trap based quantum computing devices.

\begin{figure}[tb]
	\centering
	%\begin{tabular}{ c   }
	\includegraphics[width=1.0\linewidth]{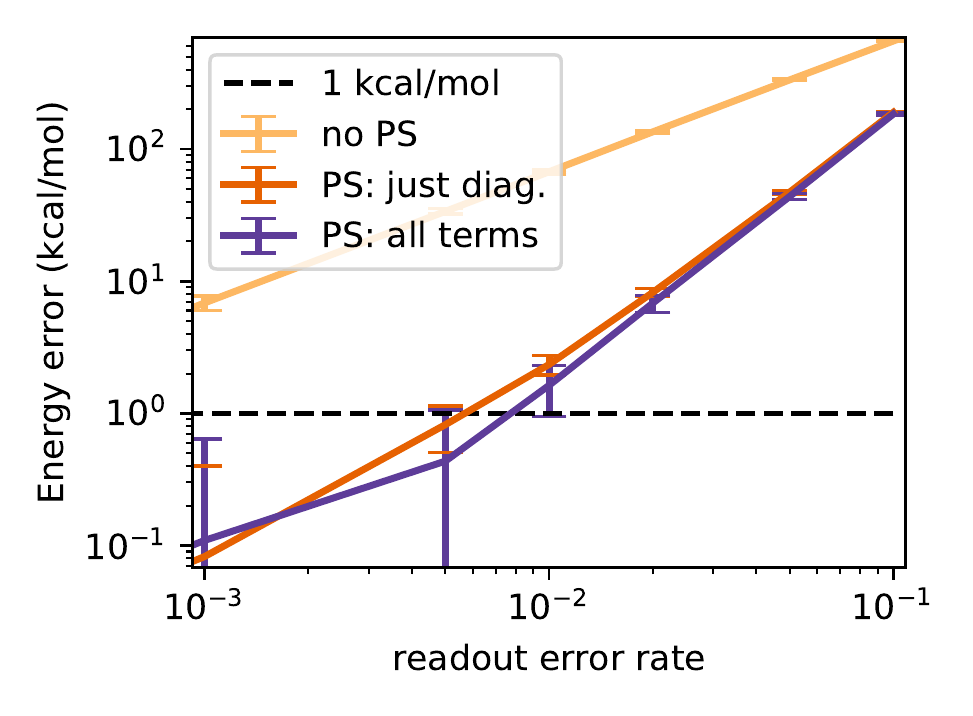}
	%\end{tabular}
	\caption{LiH pUCCD objective function estimated error in energy, as a function of readout error rate. The dashed black line corresponds to 1 kcal/mol error, known as `chemical accuracy' but used here as a reference precision level versus the readout error-free simulation result. The pUCCD circuit was executed at the optimized gate angles obtained using an exact wavefunction simulator at interatomic distance $d=1.595$ angstrom. "No PS" indicates the performance when no post-selection is applied. "PS: just diag." indicates error mitigation by post-selection was applied, filtering all terms in the Z basis only. "PS: all terms" indicates post-selection was also applied on XX and YY terms after rotating pairs of qubits to the XX+YY basis with a rotation gate as in \eqnref{rotation_gate}.}
	\label{postselectionplot}
\end{figure}

In \figref{postselectionplot} we show post-selection error mitigation results for a particular example of LiH molecule in $STO-3G$ basis set at equilibrium bond distance. With increasing noise but no error mitigation, we find relatively poor performance even with low $10^{-3}$ readout error rates. At a readout error of $10^{-2}$, doing post-selection only on the originally-diagonal terms yields a large increase in accuracy down to almost 1 kcal/mol. Next, if we also rotate all XX+YY terms to the particle basis and perform post-selection on the results, we find a further slight improvement. For an equivalent accuracy of 1 kcal/mol, the additional error mitigation of XX+YY terms allows for a $45\%$ higher readout error rate. 

In future work, we aim to find \textit{general} trends for this observation for a broader set of test cases; different molecules, geometries and basis sets. We also aim to investigate whether there is still a way to measure all  $\mathcal{O}(N^2)$ XX+YY terms simultaneously, which would effectively remove the $\mathcal{O}(N)$ scaling from the number of distinct measurement sets.

\subsection{Beyond VQE}

Regarding the computational scaling of quantum algorithms in general, beyond VQE, a similar argument as above can be proposed using the seniority-zero Hilbert space approach.
In conventional quantum phase estimation (QPE) (we refer the reader to Ref.~\cite{guzik} for a short review of the different versions of the QPE algorithms), an ancillary qubit register is added to the quantum circuit which aids in measuring eigenenergies of the Hamiltonian operator in $\mathcal{O}(1)$ measurements and $\mathcal{O}(1/\epsilon)$ circuit depth, given a trial state preparation with a significant overlap to the eigenstate of interest.
For some systems, it has been proven that the HF-state, a simple product state preparable using $\mathcal{O}(1)$ gates, may have an exponentially vanishing overlap with the true ground state.
In that case, a better state preparation scheme such as a UCC method like pUCCD may enable the subsequent use of QPE.
In either of these cases, the scaling is improved by pUCCD, because the state preparation has only $\mathcal{O}(N)$ depth and adiabatic state preparation at most $\mathcal{O}(N)$ assuming parallelizable gate operations on a linear architecture.

Further, in the phase estimation part of QPE based on Trotterized evolution of the Hamiltonian in \eqnref{bosonicHamil}, the \textit{controlled unitaries} describing the Hamiltonian evolution require a gate depth scaling at most as $\mathcal{O}(1)$ per Trotter step, while the total QPE gate depth will scale linearly in $N$.
Such a low depth can be achieved using the following strategy: considering \eqnref{qubithamil}, one part consists of operators $Z$. These operators mutually commute, and Hamiltonian evolution over these operators can be implemented as straightforward single-qubit rotations. Controlled-single qubit rotations are relatively straightforward to implement, in the context of more general controlled-unitary circuits. Another part consists of terms like $XX+YY$. We may perform a unitary transformation, on-chip, using just a fixed set of Givens-like rotations (as shown in \cite{babbush} for a Fermionic case). This effectively diagonalizes the Hamiltonian and the remaining controlled-$\exp( i\theta Z)$ operations are again efficient to implement. A similar separate diagonalization can be performed for the $ZZ$ terms. Effectively, an overall gate depth of $\mathcal{O}(N)$ is achieved with just $\mathcal{O}(1)$ controlled-single qubit rotations and the rest static circuit elements.

This is much less demanding implementation-wise than in the conventional unrestricted case with  $\mathcal{O}(N^5)$ controlled-unitary operations ($N^4$ terms in the Hamiltonian, requiring at most $N$ operations per term in nearest-neighbour gate hardware architectures) per Trotter step.
These controlled-unitary gate operations are often challenging to realize practically on quantum devices, as they involve multi-qubit interactions which are hard to implement coherently. 
In the un-restricted Hamiltonian QPE simulation, these controlled-unitaries involve at most 5-qubit interactions, whereas in the paired-electron Hamiltonian QPE, the controlled-unitary operations require just 2-qubit interactions and thus also allow for shorter decompositions into conventional two qubit gate sets. 
In particular, we note how the classical complexity reduction of FCI versus DOCI is much less practical (exponential to exponential) than the QPE equivalent reduction (polynomial to linear).

The performance enhancement is likewise expected for Kitaev's PEA and Iterative Phase Estimation methods, because the state preparation and controlled-unitary operations remain the main components contributing to the total runtime of the algorithms.

\section{Conclusions \& Outlook}
We have proposed a quantum algorithm paradigm for simulating molecular chemistry, leveraging the so-called paired-electron approximation to reduce the required quantum resources.
While conventional second-quantized Hamiltonians have $\mathcal{O}(N^3)-\mathcal{O}(N^4)$ sets of Hamiltonian terms which can be measured simultaneosly (depending on optimal sorting capabilities), the paired-electron Hamiltonian scheme has just $N^2$ terms, which can be sorted into just $3$ sets of terms which can be measured simultaneously with only single-qubit rotations required for the basis changes.
This results in a drastic improvement in the number of runs necessary for Hamiltonian averaging and thereby a reduction of the overall quantum computing time.
The NISQ experimental methods proposed here present an attractive opportunity for near-term experiments on noisy hardware due to its short circuit-depth and low variational runtime requirements.
The method can be combined with existing error mitigation protocols, and we showed an extension of a standard post-selection error mitigation method applied to this case with efficient term grouping.
Such experiments are generally interesting for practicing on near-term hardware at small scales even without directly aiming for a quantum advantage \cite{arute2020hartreefock}.
The pUCCD protocol enables efficient use of a parallelized circuit for Trotterized ansatz simulation with only native two-qubit gate requirements, such that the qubit lattice connectivity requirement is relaxed compared to previous UCC proposals.
Only a single linear chain needs to be defined across the lattice to reach the provably minimal linear gate depth shown in \cite{swapnetwork}.
Restricting the subspace to only include superpositions of closed shell electron singlet-configurations also allows for more efficient preparation of approximate eigen states for quantum phase estimation.
Also fault tolerant quantum computers devices may profit from an increase in computational accuracy enabled by spending freed-up quantum computing resources on a larger basis set, as well as a simplified implementation in terms of the required conditional-unitary circuits. 
The quantum circuit depth of the pUCCD technique scales linearly with the number of molecular orbitals (MOs) $N$. More specifically, the only required gates for the Ansatz are Givens rotations (in combination with SWAP gates on finite-connectivity hardware) allowing easier implementation on near-term quantum devices as compared to conventional UCC approaches. 
Specialized quantum hardware may implement such a two-body partial-swap/Givens rotation far more efficiently than conventional generalized universal-gate quantum computing hardware.
We also observed that the paired-electron approximation works rather well for nuclear geometries close to the equilibrium, and small diatomic molecules we tested.
It is also interesting that pUCCD comes very close to DOCI.
Implementations of classical analogs to pUCCD, such as pCCD, were only able ot achieve DOCI accuracy by including additional terms, leading to pCCD0, pECCD or ROCCSD0 ansatzes, to mention a few.
In future work, it might be worthwhile to consider the benefits of including such additional terms, particularly through the use of orbital optimization (OO). 
Ref.~\cite{Sokolov2020} showed how electron-paired type UCC methods may reach very good results when applying OO techniques to make optimal use of the imposed restrictions. 

To conclude, the molecular chemistry simulation scheme proposed here opens up new possibilities towards near-term quantum chemistry simulation on NISQ devices by delivering high accuracies with limited quantum resources. In addition, it presents a promising outlook for optimal use of future large-scale fault tolerant quantum devices. 

\section*{Contributions}
Covestro and Qu \& Co jointly conceived the project. V.E.E.\ \& C.G.\ provided the theoretical foundation and the specifics of the quantum algorithm. J.G.\ laid the classical-computational chemistry foundation and suggested to try the pair-restricted approximation. M.M.\ contributed to the conception of the work and framework. V.E.E.\ carried out the simulations, collected the data, and performed the data analysis. All authors wrote and revised the manuscript.

\appendix*
\section{Hamiltonian Matrix Elements}
We here detail how the Hamiltonian matrix elements $h_{p,q,r,s}$ are calculated from the electron-integrals, in the case of both the Fermionic as well as for the Hard-Core Bosonic hamiltonians.
The electron integrals calculated with \cite{pyscf} are stored in arrays of the form
\begin{eqnarray}
\textbf{e}_{\text{sei}}& \hspace{10pt} (N\times N)\\
\textbf{e}_{\text{tei}}& \hspace{10pt} (N\times N \times N \times N)
\end{eqnarray}
where N is the number of Molecular Orbitals (2 electrons fit in each orbitals, so 2N spin-orbitals in total).\\
The Fermionic Hamiltonian in second quantization and Born-Oppenheimer approximation is written as
\begin{equation}\label{fermionicHamilAppendix}
\hat{H} = C + \sum_{p,q} h_{p,q} \hat{a}_p^\dagger\hat{a}_q+\frac{1}{2}\sum_{p,q,r,s} h_{p,q,r,s} \hat{a}_p^\dagger\hat{a}_q^\dagger\hat{a}_r \hat{a}_s  ,
\end{equation}
where indices $\{p,q,r,s\}$ run from $1$ to $2N$, where even indices indicate alpha electrons and odd indices are beta electrons. Then
\begin{eqnarray}
h_{2i, 2j}=h_{2i+1, 2j+1}=\textbf{e}_{\text{sei}}[i,j],
\end{eqnarray}
and
\begin{eqnarray}
h_{2i, 2j, 2k,2l}&=&\textbf{e}_{\text{tei}}[i,j,k,l]\\
h_{2i+1, 2j+1, 2k+1,2l+1}&=&\textbf{e}_{\text{tei}}[i,j,k,l]\\
h_{2i, 2j+1, 2k+1,2l}&=&\textbf{e}_{\text{tei}}[i,j,k,l]\\
h_{2i+1, 2j, 2k,2l+1}&=&\textbf{e}_{\text{tei}}[i,j,k,l],
\end{eqnarray}
for indices $i$ and $j$ running from $1$ to $N$.
while the doubly-occupied only, or HCB, Hamiltonian can be written as
\begin{equation}\label{bosonicHamilAppendix}
\hat{H}_r = C + \sum_{p,q} h^{(r1)}_{p,q} \hat{b}_p^\dagger\hat{b}_q + \sum_{p\neq q} h^{(r2)}_{p,q} \hat{b}_p^\dagger\hat{b}_p \hat{b}_q^\dagger\hat{b}_q,
\end{equation}
where
\begin{eqnarray}
h^{(r1)}_{i,i}&=&2\textbf{e}_{\text{sei}}[i,i]+\textbf{e}_{\text{tei}}[i,i,i,i]\\
h^{(r1)}_{i,j}&=&\textbf{e}_{\text{tei}}[i,i,j,j] \hspace{10pt} (i\neq j)\\
h^{(r2)}_{i,j}&=&2\textbf{e}_{\text{tei}}[i,j,j,i]-\textbf{e}_{\text{tei}}[i,j,i,j] \hspace{10pt} (i\neq j)
\end{eqnarray}
for indices $i$ and $j$ running from $1$ to $N$.

%\bibliography{zsp_QC_QC}

\end{document}